\documentclass[11pt,letterpaper,notoc]{JHEP}
\newcommand{\sech}{{\rm sech}}
\newcommand{\csch}{{\rm csch}}
\newcommand{\beq}{\begin{eqnarray}}
\newcommand{\eeq}{\end{eqnarray}}

\def\ltap{\ \raise.3ex\hbox{$<$\kern-.75em\lower1ex\hbox{$\sim$}}\ }
\def\gtap{\ \raise.3ex\hbox{$>$\kern-.75em\lower1ex\hbox{$\sim$}}\ }

\def\be{\begin{equation}}
\def\ee{\end{equation}}
\newcommand{\eref}[1]{(\ref{#1})}
\newcommand{\Eref}[1]{Eq.~(\ref{#1})}

\newcommand{\Tev}{ {\rm TeV} }

\title
{Large Extra Dimensions from a Small Extra Dimension}
\author{
Z. Chacko\footnote{zchacko@phys.washington.edu}, 
Patrick J. Fox\footnote{pjfox@phys.washington.edu}, 
Ann E. Nelson\footnote{anelson@phys.washington.edu} 
\hskip 0.2cm 
and Neal Weiner\footnote{nealw@phys.washington.edu}\\
Department of Physics, Box 1560, University of Washington, 
                 Seattle, WA 98195-1560, USA
}
\preprint{UW/PT-01/16}
\abstract{Models with extra dimensions have changed our understanding 
of the hierarchy problem. In general, these models explain the
weakness of gravity 
 by diluting gravity in a large bulk volume, or by 
localizing the graviton away from the standard model. In this paper, 
we show that the warped geometries necessary for
the latter scenario can naturally induce 
the large volumes necessary for the former. We present a model
 in which a large volume is  
stabilized without supersymmetry. We comment on the phenomenology of  
this 
scenario and generalizations to additional dimensions.}

\begin{document}


%
----------------------------------------------------------------------------
\section{Introduction}
----------------------------------------------------------------------------

One of the most exciting developments in the past few years has been 
the recognition that theories with extra dimensions can radically 
change our understanding of old problems. That extra dimensions could 
be relevant for four dimensional theories is not a 
new idea, going back to the theories of Kaluza and Klein 
\cite{Kaluza:1921tu,Klein:1926fj,Klein:1926tv}. 
However, much of the recent interest has been sparked by the 
possibilities that extra dimensions offer to change our understanding 
of the hierarchy problem, which, simply put, is the question of the 
origin and stability of the large ratio of the Planck scale $M_{Pl}$ to 
the
electroweak symmetry breaking scale $M_W$. For a recent
review and more references, see ref.~\cite{Rubakov:2001kp}.

In ref. \cite{Arkani-Hamed:1998rs,Antoniadis:1998ig}, Arkani-Hamed,
Dimopoulos and Dvali (``ADD'') and Antoniadis
noted that with $n$ compact  extra dimensions,
and factorizable geometry with volume $V_n$, the effective four 
dimensional 
Planck scale $M_{Pl}$ was related to the higher-dimensional 
gravitational 
scale $M_{*}$  by the relation
\begin{equation}
        M_{Pl}^{2}=M_{*}^{2+n} V_{n}\ .
        \label{eq:addeq}
\end{equation}
They then proceeded to consider the possibility that $V_{n}$ was 
exponentially large, such that $M_{*} \sim M_{W}$. Remarkably, for 
$n\ge 2$, such a scenario was phenomenologically viable, provided the
standard model fields are confined to a 3+1 dimensional subspace---a 
``3-brane''. 
The hierarchy 
problem is recast not as a question of why $M_{W}$ is small compared 
to $M_{Pl}$, but rather, why $V_{n}$ is so large. 

In ref. \cite{Randall:1999ee}, (``RS1''), Randall and Sundrum 
considered an 
alternative possibility. Rather than focus on large, factorizable 
compact
dimensions, they noted that a non-factorizable ``warped'' product of  a 
fifth
dimension with our  four had extremely interesting
implications. Specifically, the scale factors of Poincar\'e invariant 
3-branes
embedded at different locations in 5D 
anti-de Sitter space (AdS) differ exponentially. Hence obtaining an
exponential hierarchy of scales merely required  a slice of
5D AdS  between two 3-branes. Gravity was localized to the brane with a
large warp factor (``the Planck brane''), while standard model
particles were localized to the brane with the small warp factor
(``the TeV brane''). To obtain a sufficiently large hierarchy, the size 
of the fifth dimension had to be about  $40$ in 
units of the fundamental scale.  
Goldberger and Wise demonstrated that such a dimension could 
naturally be stabilized with a bulk scalar field
\cite{Goldberger:1999uk}, providing a solution to the hierarchy
problem.
In a second paper \cite{Randall:1999vf}, (``RS2''), Randall and Sundrum 
showed that such localization of gravity could  obviate the need for
compactification of the fifth dimension. 

Models have been proposed to generate the 
exponentially large compact volumes 
of the ADD scenario  \cite{Arkani-Hamed:1999dz,Albrecht:2001cp}. 
However these  contain massless bulk scalars. Also, the
bulk cosmological constant must be extremely small. (This requirement
is in addition to the fine-tuning needed to make the effective 4-D
cosmological constant sufficiently
small.) Thus  explaining the hierarchy 
via large extra dimensions seemed to require  fine-tuning of several
parameters. Bulk 
supersymmetry can render these fine-tunings natural
\cite{Arkani-Hamed:1998kx,Antoniadis:1998ax}, 
but ideally one would not need to invoke supersymmetry 
to solve the hierarchy problem with large extra dimensions.
 Also, these models contain  light radions, with mass less than  an 
meV,
which  pose a  severe challenge for cosmology 
\cite{Arkani-Hamed:1999gq,Csaki:1999ht}. 
Given the dearth of models, it has been difficult to examine 
the validity of various ``model independent'' claims about the
required features of the stabilization mechanism, and the radion
phenomenology and cosmology.

In this paper, we will present a mechanism capable of generating
exponentially large compact dimensions without any supersymmetry. We 
take advantage of the fact that in warped six (and higher) dimensional 
geometry,  the Randall-Sundrum mechanism  can be used to generate
a hierarchy of scales. 
However, rather than using the mechanism
 to localize gravity on another brane, we will use it to generate an an 
exponentially large
volume for one or more compact dimensions. We will show that such
exponentially large volumes can be stabilized with a scalar field, 
analogous to 
the Goldberger-Wise mechanism.  We give a specific example, which is
very similar to the  six-dimensional  Randall-Sundrum type
model considered by Chacko and Nelson 
\cite{Chacko:1999eb}.  In that model gravity was localized to a 
4-brane (``the Planck brane'') with a large warp factor, while the 
standard model  resided on a different
brane with a small warp factor 
(``the TeV brane''). One dimension of the 4-brane was compact, but had
a large size in terms of the fundamental scale. 
The main  new feature of the example
considered here  is that we now place the standard
model on the same 4-brane which localizes
gravity, and there is no need for a second 4-brane. 
The standard model particles are confined to a
3-brane which is embedded in the 4-brane. Gravity is weak due to the 
presence of
a large compact dimension, and the fact that gravity spreads evenly 
over the
entire 4-brane.

The central elements of our  scenario are simple and can be generalized
to any number of compact dimensions.
\begin{enumerate}
\item We will require
a 3+$n$ brane wrapped around $n$ compact dimensions of a warped 5+$n$ 
dimensional
space-time. These $n$ dimensions  play the
role of the large extra dimensions of the ADD scenario, and we will
refer to them as ``ADD dimensions''. The warp
factors of the ADD dimensions as well as of our usual 4 depend on an
additional dimension, which we refer to as the RS dimension. 
\item Due to the presence of a non-trivial warp factor, exponentially large
hierarchies
can arise naturally. The physics which stabilizes the size of the ADD
dimensions is sensitive to these warpings. As a consequence, the volume of
the ADD space can be quite large.
\item The particles of the standard model are confined to a three 
brane, which
  is embedded in the 3+$n$ brane. Gravity, of course, lives
  everywhere, but is mostly localized to the 3+$n$ brane. Gravity 
appears weak to us  because it is diluted by
  the large ADD dimensions.
\end{enumerate}

The layout of the paper is as follows: in the next section we 
consider a toy six-dimensional model with one 4-brane. 
The requirement that the 
metric be regular everywhere  stabilizes the ADD dimension, while 
adding a bulk scalar field will naturally achieve a large volume. 

 In section \ref{sec:other}, we discuss the 
 possibilities for generalizations to 
higher dimensions. 
In section 
\ref{sec:pheno} we discuss the phenomenology of our scenario.

\section{A Simple Model}
To achieve a sufficiently large volume such that the fundamental
scale, $M_{*}$, is as low as $ \sim \Tev$,  we must realistically 
generate a large volume for at least 
two additional dimensions. For simplicity,   here we present a example 
in which an 
exponentially large volume is generated for one additional dimension. 
Generalizations to more dimensions should follow straightforwardly 
and we comment on them in section \ref{sec:other}.

The setup will be simple:
we consider the case of a six dimensional space with two extra compact
directions. We will wrap a 4-brane around one of these dimensions 
and take the space to be orbifolded across it. We take the bulk
cosmological constant to be negative (AdS). Without bulk matter, the
space  will 
want to expand to infinity. To stabilize the compact dimensions, we 
will add a massive 
bulk scalar field with a source on the brane. If the mass of this
scalar is   lighter than the fundamental scale by a factor
of a few, the setup will
be stabilized at finite but exponentially large volume.

We label a general coordinate by $x^M$ where $M$ takes values
from 0 to 3,5 and 6. The four dimensional coordinates are labeled by
$x^{\mu}$ while the two extra coordinates are $r$ and $\phi$. The
coordinate $\phi$ runs from 0 to $2\pi$. The space is
assumed to be orbifolded  about the 4-brane which is at a specific
location $r=b$ in the higher dimensional space. In addition to this 
4-brane, the sources of gravity  are a bulk cosmological constant
$\Lambda_B$, and a bulk scalar field which has a source on the brane. 
We also
assume another form of matter which is localized to the brane and whose 
effect is to make the brane tension anisotropic, such as a 
 flux \cite{Sundrum:1998ns,Arkani-Hamed:1998kx}, 
the Casimir energy of  massless fields \cite{Chacko:1999eb} or
a complex scalar field with a non-trivial winding number in the compact 
direction. 

The gravitational action for our system is given by

\begin{equation}
S_G = \int d^6 x \sqrt{-G} (2 M_*^{D-2} R - \Lambda_B)
- \sqrt{-\bar G} \delta(r-b)) \bar{\Lambda})\ .
\end{equation}
The action for the bulk scalar field is given by
\begin{equation}
S_M =  \int d^6 x \sqrt{-G} \frac{1}{2}
(-\partial_M \psi \partial^M \psi - m^2 \psi^2) +
\sqrt{-\bar{G}}H(\psi) \delta(r-b)\ .
\end{equation}
The action for the fields localized to the brane takes the form
\begin{equation}
S_W =  \int d^6 x \sqrt{-\bar{G}} \delta(r-b) L_W 
\end{equation}

We will obtain an approximate solution for the metric of this coupled 
gravity-matter system using the method of ref. \cite{Chacko:1999eb}. We 
first 
obtain a metric solution for the gravitational part of the
action alone. We then solve for the scalar field in this background as 
a
perturbation. The backreaction of the scalar field on the metric then
determines the solution of this system to leading order in the source
for the scalar field.  
 
For simplicity we will be taking
\begin{equation}
H(\psi) = \lambda \psi\ .
\end{equation}

The metric for this system has the general form
\begin{equation}
ds^2 = f(r) \eta_{\mu \nu} dx^{\mu} dx^{\nu} + s(r)d\phi^2 + dr^2.
\end{equation}
The solution for the metric from the gravitational part of the action
alone for $r<b$ 
is given by \cite{Chodos:1999zt} (see also \cite{Chacko:1999eb},
\cite{Chen:2000at, Leblond:2001xr}),
\begin{eqnarray}
f_0(r) &=& \frac{\cosh^{\frac{4}{5}} \alpha r}{\cosh^{\frac{4}{5}} 
\alpha
b}\ ,
\\
s_0(r) &=& \frac{\sinh^2 \alpha r}{\alpha^2 \cosh^{\frac{6}{5}} \alpha
r}\ .
\end{eqnarray}
Here we are normalizing $f$ to be one at the location of the brane and $\alpha$ is
defined by {\footnote{This definition of $\alpha$ differs from that in
\cite{Chacko:1999eb}.}}
\begin{equation}
\alpha^2  = -\frac{5}{16} \frac{\Lambda_B}{2M_*^{4}}\ .
\end{equation}
The corresponding solutions for $r>b$  are determined by the symmetry
condition of the orbifold. 

We will see shortly that this setup is not an extremum at finite $b$. 
Hence, we will delay our discussion of the matching conditions until 
we have included the additional scalar field and asymmetric brane tensions.
The corrections to this geometry due to the scalar field are 
parametrized
by
\begin{eqnarray}
f &=& f_0(1 + \epsilon). \\
s &=& s_0(1 + \kappa).
\end{eqnarray}
The Einstein equations linearized in $\epsilon$ and $\kappa$ take the 
form
\begin{eqnarray}
\frac{3}{2} \epsilon'' + 3 \frac{f'_{0}}{f_{0}} \epsilon' + \frac{1}{2} 
\kappa'' + \frac{s'_{0}}{2 s_{0}} \kappa' + 
\frac{3}{4} \frac{f'_{0}}{f_{0}} \kappa' 
+ \frac{3}{4} \frac{s'_{0}}{s_{0}}\epsilon' &=& \overline{T}, \\
2 \epsilon'' + 5 \frac{f'_{0}}{f_{0}} \epsilon' &=& \overline{T}, \\
3 \frac{f'_{0}}{f_{0}}\epsilon' + \frac{s'_{0}}{s_{0}}\epsilon' 
+  \frac{f'_{0}}{f_{0}} \kappa'
&=& \tilde{T},
\end{eqnarray}
where $\overline{T}$ and $\tilde{T}$ are related to the stress tensor 
 $T$ for the bulk scalar by 
\begin{eqnarray}
\overline{T}&=& \frac{{T^0}_0}{2 M_*^4 } = \frac{{T^5}_5}{2 M_*^4 },
\\
\tilde{T}&=& \frac{{T^6}_6}{2 M_*^4 }.
\end{eqnarray}
They are not independent but are constrained by energy-momentum
conservation $\nabla_M T^{MN} = 0$ which when linearized implies
\begin{equation}
\tilde{T}' = \frac{(f^2 s^{\frac{1}{2}})'}{f^2 
s^{\frac{1}{2}}}(\overline{T} - \tilde{T}).
\end{equation}
The linearized 55 equation above can be solved to give 
\begin{equation}
\epsilon' = f^{-\frac{5}{2}}\int_0^r d\rho \frac{1}{2} f^{\frac{5}{2}}
\overline{T} 
\end{equation}
Substituting for $\overline{T}$ in terms of $\tilde{T}$ in this
expression and integrating by parts this reduces to
\begin{equation}
        \epsilon'=\frac{\tilde{T}}{4 \alpha}\tanh(2 \alpha r) + 
        \frac{1}{4}\sech^{2}(\alpha r) \int_{0}^{r} d\rho \tanh^{2}(2 
\alpha 
        \rho) 
        \tilde T\ .
\end{equation}
For large $\alpha r$ keeping only the leading and subleading terms
in an expansion in $e^{-2 \alpha r}$ this further reduces to
\begin{equation}
\epsilon' = \frac{1}{4\alpha}\tilde{T} + D(r) e^{-2 \alpha r}\ ,
\end{equation}
where
\begin{equation}
D(r) = \int_0^r d\rho \tanh^2\left(2 \alpha \rho\right)\tilde{T}\ .
\label{eq:deq}
\end{equation}

Using the linearized 66 Einstein equation,  we solve for 
$\kappa'$ in the same region $\alpha r \gg 1$ 
\begin{equation}
\kappa' = \frac{1}{4\alpha}\tilde{T} - 4 D(r) e^{-2 \alpha r}\ .
\end{equation}

We can now address the stability of the setup by investigating the 
matching conditions at the brane.
Linearizing in $\epsilon$ and $\kappa$, these take the form
\begin{eqnarray}\label{match1}
&\frac{3}{2}\Delta\frac{f_0'}{f_0} + \frac{1}{2}\Delta\frac{s_0'}{s_0} 
+ 
\frac{3}{2} \Delta \epsilon' + \frac{1}{2} \Delta \kappa' = \frac{1}{2}
\lambda \psi(b) - \beta^2,& \\ \label{match2}
&2\Delta\frac{f_0'}{f_0} + 2 \Delta \epsilon' = \frac{1}{2}
\lambda \psi(b) -  \gamma^2\ ,&
\end{eqnarray}
where $\beta^2={T^*}_0^0/ 2M_*^4$ and  $\gamma^2={T^*}_5^5/ 2M_*^4$. 
Here $T^*$ is
the brane tension. 
The anisotropy between $\beta^2$ and $\gamma^2$ is due to the 
matter field localized on the wall, which we are keeping at lowest 
order in our
calculation. As discussed in ref. \cite{Chacko:1999eb},    for large
$\alpha b$ the anisotropy is an exponentially small effect.

Taking the difference of  equations \eref{match1} and \eref{match2}, 
and using the expressions
obtained earlier for $\epsilon'$ and $\kappa'$, we can find an 
equation for $D(b)$. Using the explicit forms for $f_{0}$ and $s_{0}$, 
we have
\begin{equation}
-5 D(b) e^{-2 \alpha b} = \beta^{2}-\gamma^{2}-4 \alpha\ \csch 
(2 \alpha b)\ .
\label{eq:Deq}
\end{equation}
Notice that without the scalar field, $D(b)$ vanishes.  $\beta^2-\gamma^2\sim
1/M_*^4 s^{-5/2}$ which for large $b$ is $O\left( \alpha e^{-2 \alpha
    b}\right)$.  Thus if $(\beta^{2}-\gamma^{2})/e^{-2 \alpha  b}>8 \alpha$,
then, without the scalar field, \Eref{eq:Deq} can only be satisfied (trivially)
with $b=\infty$.  The scalar field provides an extra attractive force which
stabilizes the setup.  With the scalar field present and the inequality
satisfied there is a solution for finite $b$.

It is also interesting to note that 
$D(b)$, which is a small, subleading effect, is critical in
determining $b$. This suggests that there will be a  light scalar 
mode in the lower dimensional effective theory, analogous to the
radion of the Goldberger-Wise model \cite{Goldberger:1999un}. 

To determine  $b$, we will
 obtain an approximate expression for $D(b)$. To do this
we consider the equation of motion for the bulk scalar:
\begin{equation}
- \psi'' - \left(2 \frac{f'_{0}}{f_{0}} + \frac{1}{2} 
\frac{s'_{0}}{s_{0}}\right)
 \psi' + m^2 \psi
= \lambda \delta(r-b)\ ,
\end{equation}
which, given the forms of $f_{0}$ and $s_{0}$ is just
\begin{equation}
- \psi'' - 2\alpha \coth(2 \alpha r) \psi' + m^2 \psi
= \lambda \delta(r-b)\ .
\end{equation}
While this equation is difficult to solve exactly, it simplifies 
greatly in in large and small $r$ limits. For our purposes, it will 
be sufficient to obtain approximate solutions, using the asymptotic 
forms
\begin{eqnarray}
\psi &=& A_{1} I_{0}(m r) \approx A_1 \left(1 + \frac{1}{4} m^2 r^2 +
\ldots\right) \;\;\; r\ll \frac{1}{\alpha}\ , 
\label{eq:asym1}\\
\psi &=& B_{1} e^{\sigma_1 r} +  B_2 e^{\sigma_2 r} \;\;\; 
\frac{1}{\alpha} \ll r<b\ .
\label{eq:asym2}
\end{eqnarray}
Here $\sigma_1$ and $\sigma_2$ are given by
\begin{eqnarray}
\sigma_{1} &=& -\alpha - \sqrt{\alpha^2
+ m^2},\\
\sigma_{2} &=& -\alpha + \sqrt{\alpha^2
+ m^2}\ .
\end{eqnarray}

To obtain an approximate expression for $D$ we now assume that the 
form  of \Eref{eq:asym1}, valid for $r \ll \frac{1}{\alpha}$, holds out 
to a point
$a \approx \frac{1}{\alpha}$, and that the form of \Eref{eq:asym2} 
holds from the point $a$ all the way to $b$. This will suffice
for an order of magnitude estimate of $b$. Further we will be 
interested  
in the case $m^2 \ll \alpha^2$. Then $\sigma_{1} \approx
-2 \alpha$ and $\sigma_{2} \approx 
\frac{m^2}{2 \alpha}$.
Matching values and first derivatives at $r=a=1/\alpha$, and satisfying 
the jump condition at $r=b$, we find
\begin{equation}
        A_{1} \approx \frac{\alpha \lambda}{m^{2}}e^{-b m^{2}/2\alpha}, 
\hskip 0.5in B_{1} \approx 
        \frac{(1-a \alpha) e^{2 a \alpha } \lambda}{4 \alpha}
                e^{-b m^{2} /2\alpha}, 
\hskip 0.5in B_{2} 
        \approx \frac{\alpha \lambda}{m^{2}}e^{-b m^{2}/2 \alpha }.
\end{equation}
The expression for
the stress tensor is 
\begin{equation}
\tilde{T} = \frac{1}{8 M_*^{4}} ({\psi'}^2 - m^2 \psi^2).
\end{equation}
In order that the stress tensor of the scalar field is a 
perturbation on the background metric we assume that
$\lambda \ll \alpha^3$. With these approximations, we 
find
\begin{equation}
        D(b) = \frac{\alpha^{3} \lambda^{2}}{8 M_*^{4} m^{4}}\left( e^{-\frac{b 
        m^{2}}{\alpha}}-1\right) .
\end{equation}
At this order in $m^{2}$, this result comes entirely from that large 
$r$ region in the integral of eq. (\ref{eq:deq}), justifying the 
approximations of equations (\ref{eq:asym1}, \ref{eq:asym2}).
The equation determining $D(b)$, eq. (\ref{eq:Deq}), then gives
\begin{equation}
        e^{\frac{-b m^{2}}{\alpha}}=1+\frac{8 M_*^{4} m^{4}}{5 \alpha^{3} 
        \lambda^{2}}(8 \alpha - Y_{C}),
\label{eq:DeqII}
\end{equation}
where $Y_C$ is a constant of $O(M_{*})$  parametrizing the asymmetry in the
4-brane tension.  $Y_{C}=(\beta^2-\gamma^2)/e^{-2\alpha b}$, for large $b$.  For
a solution to exist at finite $b$ with the scalar field present the anisotropy
on the brane must be large enough, i.e. $Y_C>8\alpha$.

In order to first obtain a quick estimate 
we assume that $M_*$ and $m^2/\lambda$ are both
of order one in appropriate powers of $\alpha$. This yields \cite{Chacko:1999eb}
\begin{equation}
b \approx O\left(\frac{\alpha}{m^2}\right)
\end{equation}
In this limit $\alpha b \gg 1$ since $\alpha^2 \gg m^2$, and hence the volume of the
extra dimension is large.

More generally, assuming the right-hand side of eq. (\ref{eq:DeqII})
 to be some number 
$1/Q <1$, we have 
\begin{equation}
        e^{\frac{b m^{2}}{\alpha}} \approx Q \Rightarrow e^{b \alpha} 
        \approx Q^{\alpha^{2}/m^{2}}.
\end{equation}
The radius of the large dimension is then roughly \be
r_c=V \approx  \alpha^{-1}e^{2 b \alpha /5}= \alpha^{-1} Q^{2\alpha^{2}/5 
m^{2}}
=\alpha^{-1}e^{\left({2\alpha^2\over 5 m^2} \ln Q\right)}
\ .
\ee
For $\alpha/m$ and $Q$ of order a few, this dimension is 
exponentially large, realizing the scenario of
\cite{Arkani-Hamed:1998rs} for one dimension.

\section{Modifications and generalizations}
\label{sec:other}
 A number of 
alternatives are possible to stabilize the ADD dimensions. Instead of 
using the regularity of the 
metric at the origin, we could have instead had an ``inner brane'' at
some location 
$r=a$, where $s$ and $f$ are exponentially small. The brane spacing
could be stabilized using a bulk scalar field.  The inner brane could  
lie at an orbifold fixed point of a 
compact dimension, or, we could extend the space beyond the inner brane 
to a regular origin, 
as we have done here. In the former case,  dynamics on the 
inner brane would  determine the volume of the 
ADD space,   which would naturally stabilize  the size to the inverse 
of
the (exponentially small)   scale of inner brane physics.

It should be straightforward to generalize this mechanism to more 
dimensions in order to make it  phenomenologically viable with a
TeV quantum gravity scale. The 
most straightforward generalization would be to compactify  ADD
dimensions on hyperspheres, 
in which case regularity at the origin would still be well defined,
and could be used to stabilize their size. 
Alternatively, one could employ concentric $n-$tori, and use the 
dynamics of the interior brane  to determine the overall volume of 
the setup.

\section{Phenomenology}
\label{sec:pheno}
\begin{table}
\centerline{\begin{tabular}{|l|c|c|c|c|}
\hline
 Energy\hfil  & ADD & 
RS1 & RS2 & heterotic \\ \hline
&&&&\\ [-8pt]
$E < 1/ {r_c}$&SM + 4D GR&SM + 4D GR&SM +
4D GR &SM + 4D GR\\
&+radions&+ radions& + power law corr.& +
radions\\
 &&&&\\ \hline &&&&\\ [-8pt]
$E \sim 1/ {r_c}$&SM + 4D GR&see below &$"$&SM + 4D GR\\
&  +  grav. KK modes&&&+ grav. KK modes \\
& & ($1/r_c\sim \Tev$)&&+radion \\
 &&&&\\\hline &&&&\\ [-8pt]
$ 1/ {r_c}< E $&SM   &NA&$"$&SM + (4+$n$)D
GR\\
$ \ltap \Tev$& + (4+$n$)D GR  &&&+power law corr.\\
&&&&+ radion\\
 &&&&\\ \hline &&&&\\ [-8pt]
$E\sim\Tev$&strong quant. &SM+ 4D GR&$"$ &strong quant.\\
& grav.&+ grav. KK modes& &grav. \\ 
&&&&\\ \hline &&&&\\ [-8pt]
$E \gtap \Tev$  &$"$&5D  AdS  quant. grav. &$"$&$"$            \\
&&+ 4D GR&&\\
&&(+ other 5D fields)&&\\
&&&&\\ \hline &&&&\\ [-8pt]
$E \sim M_{Pl}$  &$"$?&strong quant.&strong quant.&$"$?            \\ 
 &&gravity&gravity&            \\ 
&&&&\\ \hline
\end{tabular}}
\vskip .2 in 
\caption{Summary of effective theories at various scales in the ADD, RS
and  heterotic extra dimensional scenarios.}
%
\end{table}

The phenomenology of our scenario is very similar to that of the  ADD
proposal \cite{Arkani-Hamed:1999dz}, however the presence of the RS 
dimension leads to a some
interesting distinctions. The relation between the Planck scale and
$M_*$ is  approximately
\be
\label{eq:heteq}
        M_{Pl}^{2}=\left({M_*^3\over\alpha}\right)M_{*}^{n} V_{n}\ ,
\ee
which is quite similar to \Eref{eq:addeq}, but with an additional
factor of $M_*/\alpha$ from the RS dimension. Note that \eref{eq:addeq}
would give the same result for $n$ large dimensions and one  smaller
dimension of size $1/ \alpha$.  However  an important difference is 
that the
gravitational  modes of the RS
dimension cannot be neglected in the effective theory below the scale
$1/\alpha$, as those whose wave functions are small in the large warp
factor region  are very  light, with masses of order $1/r_c$. These
light modes
are very  weakly coupled on the brane,   but their effects will show
up as  nearly power law corrections to the $n$
dimensional gravitational potential,  as  in the  RS2 model.
In principle the  modes of both the RS  and ADD dimensions could show 
up in collider
searches for higher dimensional graviton emission at high energies. 
The radion for the RS dimension will be somewhat lighter than the
scale $1/r_c$ and also may produce observable deviations from $1/r^2$
gravitational forces at long distances, although its wavefunction
on the brane is  small.

Because our scenario shares  interesting  features  of  both the ADD 
and RS extra
dimensional models, we refer to it as ``heterotic''\footnote{This
term, meaning `` possessing hybrid vigor'' was introduced into physics
in ref.~ \cite{Gross:1985dd}.}.
  
In Table 1 we summarize and compare the main features of  the 
phenomenology of the heterotic scenario with those of
ADD, RS1, and RS2, by describing the relevant effective theories as a 
function 
of scale. We use the abbreviations  ``SM'' for Standard Model, ``$n$D
GR'' for $n$-dimensional General Relativity, and ``KK'' for the
(Kaluza-Klein) higher dimensional modes of bulk fields.

\section{Summary}

In this paper, we have offered a mechanism to stabilize  exponentially 
large dimensions, using the ideas of Randall-Sundrum and
Goldberger-Wise for generating and stabilizing exponentially different 
scales in warped geometry.
It is amusing that    warped geometry, 
the central element in the RS1 and RS2 scenarios  
\cite{Randall:1999ee,Randall:1999vf}, in
combination with
additional compact dimensions, can naturally yield a scenario  very
similar to ADD.

A great deal remains to be studied, including  generalizations to 
higher 
dimensions, models with interior branes, the effects of our 3-brane on
the geometry, the mass(es) and coupling(s) of
the radion(s), and cosmology. 
\vskip 0.25in
{\noindent \bf Acknowledgements} 
\vskip 0.15in
\noindent This work was partially supported by the DOE
under contract DE-FGO3-96-ER40956. 
\bibliography{exex}

\providecommand{\href}[2]{#2}\begingroup\raggedright\begin{thebibliography}{10}

\bibitem{Kaluza:1921tu}
T.~Kaluza, {\it On the problem of unity in physics},  {\em Sitzungsber. Preuss.
  Akad. Wiss. Berlin (Math. Phys. )} {\bf K1} (1921) 966--972.

\bibitem{Klein:1926fj}
O.~Klein, {\it The atomicity of electricity as a quantum theory law},  {\em
  Nature} {\bf 118} (1926) 516.

\bibitem{Klein:1926tv}
O.~Klein, {\it Quantum theory and five-dimensional relativity},  {\em Z. Phys.}
  {\bf 37} (1926) 895--906.

\bibitem{Rubakov:2001kp}
V.~A. Rubakov, {\it Large and infinite extra dimensions: An introduction},
  \href{http://xxx.lanl.gov/abs/hep-ph/0104152}{{\tt hep-ph/0104152}}.

\bibitem{Arkani-Hamed:1998rs}
N.~Arkani-Hamed, S.~Dimopoulos, and G.~Dvali, {\it The hierarchy problem and
  new dimensions at a millimeter},  {\em Phys. Lett.} {\bf B429} (1998) 263,
  [\href{http://xxx.lanl.gov/abs/hep-ph/9803315}{{\tt hep-ph/9803315}}].

\bibitem{Antoniadis:1998ig}
I.~Antoniadis, N.~Arkani-Hamed, S.~Dimopoulos, and G.~Dvali, {\it New
  dimensions at a millimeter to a fermi and superstrings at a tev},  {\em Phys.
  Lett.} {\bf B436} (1998) 257,
  [\href{http://xxx.lanl.gov/abs/hep-ph/9804398}{{\tt hep-ph/9804398}}].

\bibitem{Randall:1999ee}
L.~Randall and R.~Sundrum, {\it A large mass hierarchy from a small extra
  dimension},  \href{http://xxx.lanl.gov/abs/hep-ph/9905221}{{\tt
  hep-ph/9905221}}.

\bibitem{Goldberger:1999uk}
W.~D. Goldberger and M.~B. Wise, {\it Modulus stabilization with bulk fields},
  {\em Phys. Rev. Lett.} {\bf 83} (1999) 4922--4925,
  [\href{http://xxx.lanl.gov/abs/hep-ph/9907447}{{\tt hep-ph/9907447}}].

\bibitem{Randall:1999vf}
L.~Randall and R.~Sundrum, {\it An alternative to compactification},
  \href{http://xxx.lanl.gov/abs/hep-th/9906064}{{\tt hep-th/9906064}}.

\bibitem{Arkani-Hamed:1999dz}
N.~Arkani-Hamed, L.~Hall, D.~Smith, and N.~Weiner, {\it Solving the hierarchy
  problem with exponentially large dimensions},  {\em Phys. Rev.} {\bf D62}
  (2000) 105002, [\href{http://xxx.lanl.gov/abs/hep-ph/9912453}{{\tt
  hep-ph/9912453}}].

\bibitem{Albrecht:2001cp}
A.~Albrecht, C.~P. Burgess, F.~Ravndal, and C.~Skordis, {\it Exponentially
  large extra dimensions},  \href{http://xxx.lanl.gov/abs/hep-th/0105261}{{\tt
  hep-th/0105261}}.

\bibitem{Arkani-Hamed:1998kx}
N.~Arkani-Hamed, S.~Dimopoulos, and J.~March-Russell, {\it Stabilization of
  sub-millimeter dimensions: The new guise of the hierarchy problem},  {\em
  Phys. Rev.} {\bf D63} (2001) 064020,
  [\href{http://xxx.lanl.gov/abs/hep-th/9809124}{{\tt hep-th/9809124}}].

\bibitem{Antoniadis:1998ax}
I.~Antoniadis and C.~Bachas, {\it Branes and the gauge hierarchy},  {\em Phys.
  Lett.} {\bf B450} (1999) 83--91,
  [\href{http://xxx.lanl.gov/abs/hep-th/9812093}{{\tt hep-th/9812093}}].

\bibitem{Arkani-Hamed:1999gq}
N.~Arkani-Hamed, S.~Dimopoulos, N.~Kaloper, and J.~March-Russell, {\it Rapid
  asymmetric inflation and early cosmology in theories with sub-millimeter
  dimensions},  {\em Nucl. Phys.} {\bf B567} (2000) 189--228,
  [\href{http://xxx.lanl.gov/abs/hep-ph/9903224}{{\tt hep-ph/9903224}}].

\bibitem{Csaki:1999ht}
C.~Csaki, M.~Graesser, and J.~Terning, {\it Late inflation and the moduli
  problem of sub-millimeter dimensions},  {\em Phys. Lett.} {\bf B456} (1999)
  16--21, [\href{http://xxx.lanl.gov/abs/hep-ph/9903319}{{\tt
  hep-ph/9903319}}].

\bibitem{Chacko:1999eb}
Z.~Chacko and A.~E. Nelson, {\it A solution to the hierarchy problem with an
  infinitely large extra dimension and moduli stabilization},  {\em Phys. Rev.}
  {\bf D62} (2000) 085006, [\href{http://xxx.lanl.gov/abs/hep-th/9912186}{{\tt
  hep-th/9912186}}].

\bibitem{Sundrum:1998ns}
R.~Sundrum, {\it Compactification for a three-brane universe},  {\em Phys.
  Rev.} {\bf D59} (1999) 085010,
  [\href{http://xxx.lanl.gov/abs/hep-ph/9807348}{{\tt hep-ph/9807348}}].

\bibitem{Chodos:1999zt}
A.~Chodos and E.~Poppitz, {\it Warp factors and extended sources in two
  transverse dimensions},  {\em Phys. Lett.} {\bf B471} (1999) 119--127,
  [\href{http://xxx.lanl.gov/abs/hep-th/9909199}{{\tt hep-th/9909199}}].

\bibitem{Chen:2000at}
J.-W. Chen, M.~A. Luty, and E.~Ponton, {\it A critical cosmological constant
  from millimeter extra dimensions},  {\em JHEP} {\bf 09} (2000) 012,
  [\href{http://xxx.lanl.gov/abs/hep-th/0003067}{{\tt hep-th/0003067}}].

\bibitem{Leblond:2001xr}
F.~Leblond, R.~C. Myers, and D.~J. Winters, {\it Consistency conditions for
  brane worlds in arbitrary dimensions},
  \href{http://xxx.lanl.gov/abs/hep-th/0106140}{{\tt hep-th/0106140}}.

\bibitem{Goldberger:1999un}
W.~D. Goldberger and M.~B. Wise, {\it Phenomenology of a stabilized modulus},
  {\em Phys. Lett.} {\bf B475} (2000) 275--279,
  [\href{http://xxx.lanl.gov/abs/hep-ph/9911457}{{\tt hep-ph/9911457}}].

\bibitem{Gross:1985dd}
D.~J. Gross, J.~A. Harvey, E.~Martinec, and R.~Rohm, {\it The heterotic
  string},  {\em Phys. Rev. Lett.} {\bf 54} (1985) 502--505.

\end{thebibliography}\endgroup
\bibliographystyle{JHEP}
\end{document}